# Meta-Structural properties in *Collective Behaviours* (2)

*Gianfranco Minati*
*Italian Systems Society, Via Pellegrino Rossi 42*
*Milan, I-20161, Italy*
*gianfranco.minati@AIRS.it*

*Ignazio Licata*
*ISEM, Institute for Scientific Methodology, via Ugo La Malfa 153*
*Palermo, I-90146, Italy*
*Ignazio.licata@ejtp.info*

Abstract - In this contribution we consider Collective Behaviours as *coherent sequences of spatial configurations adopted by agents interacting through corresponding different structures over time*. The approach relates to the one adopted time ago when introducing the concept of Multiple Systems arising when each composing element can play simultaneously or sequentially different roles. In this case each element simultaneously or sequentially belongs to different systems. Multiple structures over time and their sequences are considered as Meta-Structures establishing coherent sequences of different Systems. They are intended as *coherent* when acquiring emergent properties. Such coherence is considered here as being represented by the values adopted by suitable mesoscopic variables and their properties, i.e., meta-structural properties, allowing the researcher to a) *Recognise* a phenomenon as emergent; b) *Induce* emergence of collective behaviour in populations of elements collectively interacting; c) *Act* on collective emergent phenomena with the purpose to *change, regulate* and *maintain* acquired properties and d) *Merge* different collective emergent phenomena. We introduce a formal tool, i.e., *the mesoscopic general vector* to represent the adoption, over time, of mesoscopic properties by collectively interacting elements. We mention future experimental lines of activities, future lines of research and possible applications.

## 1. Introduction

In this paper we present a research project aiming at introducing and experimenting a new semi-classical approach (Minati 2008, 2009) to model coherence as related to the emergence of acquired properties in processes of collective behaviour. The approach is to be considered as a *semi-classical* one because of two main reasons. On one hand it may be considered as *classical* since it is based on considering possible to separate, distinguish, and measure components establishing phenomena of collective behaviour, non valid assumption, for instance, when dealing with quantum approaches. Furthermore interactions are considered to be explicitly represented, e.g. analytically, or modelled by using approaches based, for instance, on Cellular Automata. Phenomena where this classical view has been applied relate, for instance, to model swarms, flocks, and traffic. We will refer to such kind of collective behaviours as Spatial Collective Behaviours (SCBs) where metrical and topological assumptions are possible in contrast with quantum approaches using Spontaneous Symmetry Breaking (SSB) mechanism in Quantum Field Theory (QFT). In the later case assumptions adopted in our semi-classical approach are less efficacious when considering, for instance, quasi-particles and in quantum phenomena such as superconductivity and superfluidity. On the other hand it may be regarded as non-completely *classical* since it is based on considering the dynamics of a complex system, e.g., collective behaviour, given by simultaneous and subsequent Multiple Systems (MSs) and Collective Beings (CBs), see Section 2, established by same elements interacting in different ways as introduced in (Minati and Pessa 2006) where, however, models were originally based on classical ergodicity (Minati 2002). The new approach considered here is based on correspondent mesoscopic variables, cognitively constructed by the observer or identified as having statistical significance, such as number of elements having the *same* distance from the nearest neighbour, the *same* speed at a given point in time, the *same* direction, the *same* altitude at a given point in time, and the *same* topological position at a given point in time, where measures are considered at a suitable threshold. Moreover the approach introduced does not model phenomena of collective behaviour only by the values assumed by mesoscopic variables, but considering the *properties* of sets of values assumed by mesoscopic variables over time, such as statistical, periodicity and quasi-periodicity and named here as meta-structural properties. The possible extension of this approach to SCBs assumed as not suitably modelled by using MSs or CBs as well as to non-spatial collective behaviours requires the adoption of different mesoscopic variables, as introduced below, whereas meta-structural properties should, in any case, represent the coherence deriving from processes of emergence. The point we have to focus on here is that in many systems it is the dynamics itself (i.e., the interactions coming into play) which individuates the most suitable variables for a description. In the case of many mesoscopic systems of the "middle way" (Laughlin *et al.* 2000) this is impossible even in principle because the internal and external interactions in the system and the most part of what is considered as its "identity" changes radically. The reason to introduce this new approach having possible validity for classical representations and the related research project based on the abovementioned assumptions and representations, see Sections 4, 5 and 6, is to allow modelling collective behaviour phenomena so as to make tools available to researches to act on them, for instance, to *induce, vary, regulate maintain* and *merge* emergence of collective behaviours and their acquired properties through non-invasive or prescriptive actions, i.e., by prescribing meta-structural properties. So the approach suits when there is not a dynamics univocally defined by a Hamiltonian and, in particular, when we are far from peculiar quantum states of a system, these are what we have referred to as radical emergence situations, greatly different from the cases taken into consideration by Synergetics or Spontaneous Symmetry Breaking in QFT (Licata 2010a, 2010b).

The very first step of the project relates to the search of meta-structural properties within a simulated process of emergence such as flocks of boids and by taking into considering both metrical and topological interaction rules. This is the *minimal* conceptual framework to carry out modelling by the mesoscopic variables and searching for meta-structural properties. The outcomes and knowledge acquired from this very first experimental case is considered useful not yet to *validate* but to allow for subsequent experimental activities where values taken on by microscopic variables are available to consider mesoscopic variables and search for meta-structural properties in non-simulated cases where all microscopic information is available, such as industrial districts, markets, and swarm intelligence in plant roots when considering their electrical activity.

## 2. Concepts introduced by Multiple Systems and Collective Beings

As previously said we briefly present the related and anticipatory approach constituted by MSs and CBs already introduced in the literature (Minati and Pessa 2006). The approach is based on identifying systems having components belonging to more than a single system. Which is to say that the same components can play, both simultaneously and at different times, several roles when interacting with other components and give rise to the emerging of different systems. In this case we focus on roles allowing to model the acquired coherence by using ergodicity.

### 2.1 Multiple Systems

We recall that a Multiple System (MS) is a coherent set of simultaneous or successive systems modelled by the observer and established by the same elements interacting in different ways, i.e., having multiple roles simultaneously or at different times (Minati 2002). Examples of MSs are electrical networks composed by interconnection of electrical elements such as resistors, inductors, capacitors, sensors and actuators where different systems play different roles or networked interacting computer systems performing cooperative tasks over the Internet.

However, the reader may ask how can

- An electronic component contemporarily be component of different electronic systems?

- An organ, a cell, and a molecule be part of different biological systems, such as organisms, at the same time?

The point to be considered to understand this possibility is that different components may be simultaneously part of different systems when they play *different independent roles*.

With reference to the above listed cases we should consider that

- The output of an electronic component may be also a source of information for a security monitoring system;
- The reaction of a cell, while considering a component of a living system, may also be a source of information when using diagnostic techniques.

Even in case of multiple virtual roles we have Multiple Systems, such as when computer programs *share* involvement in different systems. That is when a computer system, while multitasking or running different work sessions, is also part of networks. In this case the component has a shared, simultaneous, participation in different systems.

Moreover, the study of MS has also considered *interchangeability* between interacting elements to model emergent behaviour as ergodic. It is well-known that ergodicity is a refined recurrence property of statistical systems. In our approach this word has a parallel and different meaning. With ergodic we mean the possibility of individuating kinds of regularity which define coherence in simultaneous or successive configurations of elements establishing systems (Minati 2002). Coherence corresponds to the acquisition of an emergent property by the MS. We remind that the *same* system can be *both* ergodic and non-ergodic depending upon the time scale of the observer, as in polymers, and temporarily ergodic as well. Besides it is possible to consider degrees and indexes of ergodicity as introduced in Section 5.2. A visual example is presented in Figure 1.

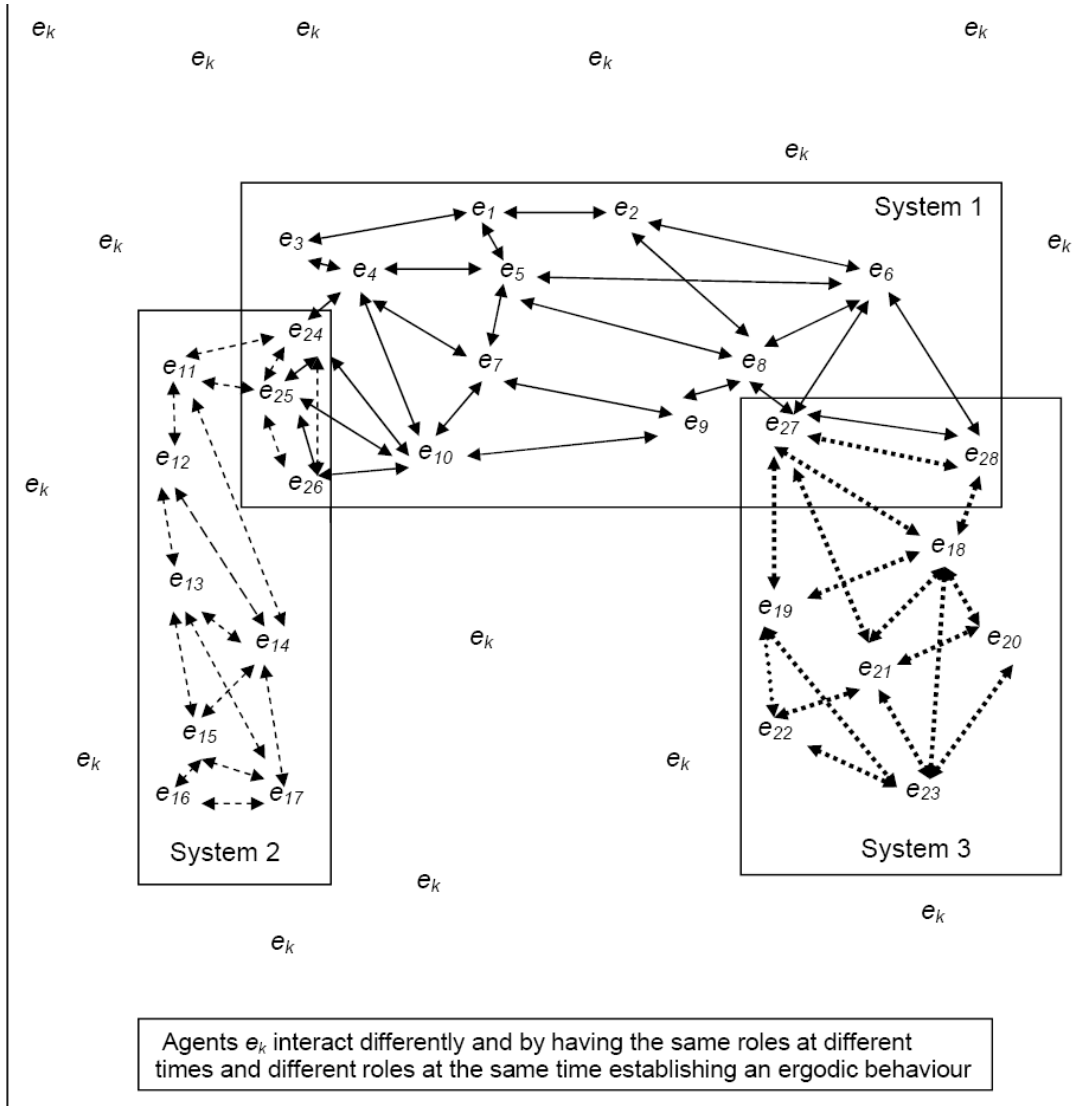

Figure 1: Visual, schematic example of flock as SCB modelled as Multiple Systems and Collective Beings of elements $e_k$

At this point we mention the possibility to consider Multiple Systems as dynamic clusters and synchronisation as the source of their coherence (see, for instance, Mikhailov and Calenbuhr 2002). For instance the case of populations of *interacting clocks*, whose internal cyclic dynamics can be given by

$$\dot{\Phi} = \omega \qquad\qquad (1)$$

where:

- $\Phi$ is the phase,
- $\omega$ is the frequency.

This is, for instance, the case of large populations of fireflies when synchronised generating a large amplitude periodic signals. It is possible to find the equation of the *Synchronisation Function* between them (Mikhailov and Calenbuhr 2002, p. 127).

Another approach (Mikhailov and Calenbuhr 2002, p. 155) is based on considering the dynamical law describing the time evolution of a generic unit represented by a *logistic map* able to represent realistically population dynamics:

$$f(x)=1-\alpha x^2 \qquad (2)$$

where:

- $x$ is the number of elements,
- $\alpha$ is a suitable control parameter.

It is possible to consider the system of $N$ *globally coupled logistic maps* to study their possible mutual synchronisation during interaction:

$$X_i(t+1)=(1-\varepsilon)\,f(x_i(t)) + \frac{\varepsilon}{N}\sum_{j=1}^{N} f(x_j(t)) \qquad (3)$$

where:

$i = 1,2, ..., N$ denotes the single logistic maps $f(x) = (1-\alpha x^2)$

$f(x)$ is given by equation $f(x)=1-\alpha x$

$\alpha$ is a control parameter for the logistic map. It displays chaotic behaviour when $\alpha > 1.401...$

$\varepsilon$ the coupling strength.

Dynamic clustering is intended as establishment of coherently operating groups. In this case "Dynamic clustering is observed inside the interval $0.32 > \varepsilon > 0.075$. Mutual synchronisation of the entire ensemble (the 'coherent phase') develops for $\varepsilon > 0.32$. If $\alpha = 1.8$, dynamical clustering takes places inside the interval $0.37\,\varepsilon > 0.14$, and so on." (Mikhailov and Calenbuhr 2002, p. 157).

From the same source (Mikhailov and Calenbuhr 2002) it is possible to find approaches for clusters and synchronisation in Dynamic Networks formed by populations of interconnected elements with simple internal dynamics. Here, the case is considered where elements consist of $N$ *identical logistic maps*: "The pattern of connections in the network is specified by a random graph with the adjacency matrix $T_{ij}$ which is obtained by independently generating any possible connection with a fixed possibility $v$." (Mikhailov and Calenbuhr 2002, p. 250).

The collective dynamics of the network is given by the equation:

$$X_i(t+1)= \left[1- [\varepsilon\,/\,v(N-1)]\sum_{j=1}^{N} T_{ij}\right]\,f(x_i(t))\ +\ \varepsilon\,/\,[\,v(N-1)]\sum_{j=1}^{N} T_{ij}\ f(x_i(t)) \qquad (4)$$

where symbols are as specified above. When $v = 1$ equation (4) coincides with (3) describing the collective dynamics of $N$ *globally coupled logistic maps.*

Numerical simulations have been performed (Manrubia and Mikhailov 1999) and showed "...that, when the coupling straight $\varepsilon$ is gradually increased, these networks experience dynamical clustering and synchronisation." (Mikhailov and Calenbuhr 2002, p. 250).

However, the limited effectiveness of these models is related to the fact that they are mainly based on global and fixed interconnections allowing for the occurrence of the phenomena of emergence as clustering and synchronisation. In our approach, instead, we consider clustering as given by cognitive states minimising energy of the neuronal phase space of the observer (Edelman and Tononi 2000). The question which meta-structures approach originates from equally pertains to cognition (observer) and physics (observed). Till when in a complex inter-relation of change processes does the observer individuate the same system (Licata and Minati 2010)? Since the very beginning the observer is called to make his choices on what and how to observe. That provides significant information on the observer's cognitive features, and the meta-structures search becomes thus aimed not only to the emergence characteristics and the change in a system regarded as "objective" -according to the classical model of observer-, but to the intrinsic, relational processuality between the observer and the observed. In this way, meta-structures are at the same a disposal to detect emergence and processuality and an important tool to study the cognitive relations put into action during observation. In other words, observing and identifying a whole as a "system" indicate a sort of cognitive homeostasis between the observer and the observed. In this sense, meta-structures are part of a future general observer-observed theory (see for instance, Heylighen 1990).

## 2.2 Collective Beings

We have a different situation when components are endowed with a cognitive system, having then *memory* of the different roles. This implies that they have a cognitive model appropriate for each role. They are autonomous agents able to select the cognitive model to be used, being each model corresponding to the system which they belong to.

Collective Beings (CBs) are particular cases of MSs when elements are autonomous agents all with the *same* cognitive system and which may *decide*, within their physical and cognitive limits, their way of interacting. When considering human CBs, these are established by using *different cognitive models*, examples are cases where elements may *simultaneously* belong to different systems (e.g., components of families, workplaces, traffic systems, consumers, and mobile phone networks) or *dynamically*, i.e., at different times, giving rise to different systems, such as temporary communities (e.g., audiences, queues, and passengers on an airplane).

*Sequences of states adopted by corresponding sequences of various single systems* established over time by the same elements interacting in different ways, i.e., having variable structures, see Section 3, establish MSs acquiring, for instance, the emergent property of a black-out in electricity networks or *coherent CBs* acquiring, for instance, the emergent properties of blocking traffic, over- or under-selling and congestion of phone lines. Their coherence is considered as being due to ergodic multiple and interchangeable roles, such as speed, altitude, distance, topological position (Ballerini *et al.* 2008) in a flock of boids. Roles considered by the observer to detect and model such coherence are not arbitrarily determined, but based on cognitive *Gestalt continuity*, extensions or replications of conceptual categories used to model normal non-collective behaviours. Creativity is intended here as *cognitive design* of mesoscopic variables, see (Licata and Minati 2010). This is why, when modelling collective behaviours, such as those of swarms and flocks, we consider, for instance, speed, altitude, phase, direction, density, and topological properties rather than age, colour, sex and weight of elements. We mention, in this regard, more specific approaches to model creativity as *the non-algorithmic procedure consisting of a jump from one model to another* (Arecchi 2011) and as in the *theory of logical openness* (Minati *et al.* 1998, Licata 2008) and the *Dynamic Usage of Models (DYSAM)* (Minati and Pessa 2006, pp. 64-75).

This research project has not the purpose of limiting the modelling of the acquisition of coherence to ergodicity, but to generalise by using Meta-Structures. We may consider, for instance, at an appropriate scale and over a discrete period of time $t_{i=1,s}$, a sequence of $m$ boids $e_{k,\ k=1,m}$ establishing a collective behaviour such a flock to be modelled as Collective Being where elements are considered to play multiple roles, such as topological positions, e.g., belong to the surface and the central area, assume the *same* distance from the nearest neighbour, the *same* speed, the *same* direction, the *same* altitude at a given point in time, where measures are considered at a suitable threshold.

## 3. The new research assumptions
In this section we present the new assumptions used for the research project.

### 3.1 Structures

While *organization* is intended to deal with networks of relations with undefined parameters, *structure deals with* networks of relations having well-defined parameters.

Structure is thus intended as an organization applied to specific cases, for instance, to components possessing characteristics and properties to which the general organizational schema is applied. More generally an *abstract structure* is given by a set of coherent relationships, degrees of freedom, rules of interaction, and properties possessed by elements. In this conceptual framework a structure of a system is considered here given by rules of interaction, constraints and suitable parameters imposed by the designer or assumed by the modeller, through which components are allowed to interact. Examples in engineering are given by:

- Mechanical devices whose components have very well-defined degrees of freedom for interacting and establishing a mechanism having an acquired property, such as a machine or an engine;
- Electronic circuits where components, when powered on, may only interact as allowed by the connections of the electronic circuit.

## 3.2 Dynamic Structures

Dynamics of systems are usually considered as properly represented on the basis of the sets of values the variables take on, e.g., microscopic or macroscopic, changing over time by following prefixed rules. In this case the dynamics are given by the interaction itself over time and not by the changing of the ways in which interaction occurs. Fixed rules of interaction constitute and represent the structure of the system. As introduced in Section 3.4 phase-transitions, for instance, are considered here as corresponding to the acquisition of, or a change in, structure. See Section 4.6 for the concept of virtual structure in collective behaviours.

## 3.3 Structures possessed by Multiple Systems

As previously reminded sets of simultaneous or successive systems established by the same elements interacting in different ways, i.e., having multiple roles simultaneously or at different times, establish MSs. Composing elements are assumed to take on the same roles at different times and different roles at the same time establishing an ergodic behaviour. When in MSs we consider ergodic behaviour reference is made, for instance, to classes of distances, speeds, altitudes and directions. In MSs elements can migrate from one class to another one and simultaneously belong to more than one. Coherence is given by their ergodic behaviour, i.e., interchangeability of roles intended as belonging to classes, as in (Minati and Pessa 2006, pp. 291-313).

In the new approach here introduced we consider the *number* of components belonging to a class per each instant. This is considering values assumed over time by mesoscopic variables as corresponding to the classes introduced above, see Section 4.2. This is where the term *Meta-Structure* originates, see Section 5.

The values which mesoscopic variables take on are intended to *represent the ways by which elements interact*. For instance, we may consider general rules of interaction valid at each temporal step (Reynolds 1987, 1999):

- alignment: elements must point towards the average motion direction of locally adjacent components;
- separation: elements must avoid the crowding of locally adjacent components;
- cohesion: elements must point towards the average position of locally adjacent components.

Elements may interact by following a general rule having specific parameters like:

- min distance $> d_1$;
- max distance $< D_1$;
- distances always change along time;
- agents may have different directions among them, but with angle $< \alpha$.

It is then possible to adopt different ranges of values specifying different, alternatives classes of minimum and maximum distance, and direction.

Besides different degrees of freedom may simultaneously and at different temporal steps apply to different elements, like:

- Classes of speed for elements;
- Classes of altitude for elements belonging, for instance, to flocks and swarms;
- Classes of topological distances and topological positions such as belonging to the upper, lower, left, right and centre areas of the surface of the SCB.

Now the question is how this sequence of structures can generate a SCB?

It is possible at this point to introduce the following propositions to be demonstrated at the accomplishment of the project introduced later.

*Proposition 1.*

*Any ergodic Multiple System possesses meta-structural properties*

Ergodicity *is* a meta-structural property when related to mesoscopic variables such as in the case of Multiple Systems. Meta-structures are represented by multiple, non-equivalent meta-structural properties, just some of them are ergodic in a proper sense. Therefore

*Proposition 2.*

*Any SCB possesses meta-structural properties, but it is not necessarily ergodic*

3.4 Coherence of sequences of structures

The concept of coherence possesses several distinct disciplinary meanings. For instance, in physics the coherence of two waves relates to their constant relative phase. Interference occurs when two waves superpose each other giving rise to a resultant wave of different amplitude. It is also possible to consider *self-coherence* when the second wave is not a separate one, but the first wave at a different time or position. In this case, the measure of co-relation is the *self-correlation* function.

Examples of other disciplinary meanings include usages in philosophy when considering the non-contradictoriness of concepts, in cognitive science when considering cognitive states, and in linguistics with reference to semantics.

In Systemics, we consider coherence the dynamic establishment and maintenance of the *same* systemic properties *continuously* set up by interacting components, e.g., an electronic device acquiring a property when powered on, leading to interactions among the component elements. The system *degenerates* into its components as soon as it is powered off, i.e., its components cease to interact. In this case, the fixed structure of the system is the source of coherence.

The reference to *same system property* is constructivistically observer-dependent since an observer provided with a proper cognitive system able to act at a suitable scalarity and level of description is required. With proper cognitive system we mean an observer possessing memory, ability to perform logical inferences, making representations and processing emotions.

A different case occurs when there is a *change in the system structure* such as in phase transitions. This change may be understood as a transition between two different coherences, i.e., phases.

We will consider here *coherence* as the property of collectively interacting elements (Aarseth *et al.* 2008) to acquire emergent properties. As mentioned below in points a), b) and c) and introduced above, this coherence is considered here as due to suitable *changes in structures*, i.e., *ways* in which elements interact and give rise to *sequences* of configurations composed of the same elements such as for MSs mentioned in Section 2. In other words, this means to consider emergent properties not as dynamical aspects of the *same* system, but as given by a sequence of configurations *of states adopted by various subsequent single systems*. This is not a mere sophism. In this case, we must ask what "the same system" means! A reductionist approach will solve such an expression in terms of constituent elements, but it is meaningless in situations where the interactions change as the identities of the constituents themselves also do, such as in many biological and social processes. It is more useful to focus on emergence as defining the coherence of a new structure at a given level of description. That is why complex systems of this kind cannot be described by a single formal model (Minati *et al.* 1998, Licata 2008)

Changes in structures producing multiple, dynamic, local, different regularities are considered here as being suitably represented by Meta-Structural properties introduced in Section 5.1. If collective interaction establishes a collective entity detected by the observer, at a suitable level of description, having properties which the component elements do not possess, we may distinguish between processes of:

(1) Phase-transitions considered here as corresponding to the acquisition of, or a change in, structure, as for first-order phase-transitions, e.g., water-ice-vapour. It is well-known that some processes of change, such as *second-order phase transitions*, consist of an internal rearrangement of the *system structure*, occurring at the same time at all points within it. In other words, the transition occurs because the conditions necessary for the stable existence of the structure corresponding to the initial phase *cease to be valid* and a new stable structure replaces it, see Sections 3.1, 3.2, 3.3. Examples are transitions from a paramagnetic to a ferromagnetic state, the occurrence of superconductivity and superfluidity, and order-disorder transitions in some kinds of crystals. In these situations there are very complicated transient dynamics where classical and quantum aspects mix (Parisi 1992, Sewell 2002).

(2) Self-organisation processes considered here as corresponding to continuous but *stable*, for instance, periodic, quasi-periodic (Hemmingsson and Peng 1994) and predictable, variability in the acquisition of new structures, as for Bènard rolls, structures formed in the Belousov-Zhabotinsky reaction, swarms having repetitive behaviour, and dissipative structures such as whirlpools in the absence of any internal or external fluctuations. Stability of variability, e.g., periodicity, corresponds to stability of the acquired property;

(3) Emergence considered here is that corresponding to the continuous, irregular and unpredictable acquisition of shapes, which become new *coherent* structures through the observer choice of a suitable cognitive model at a specific level of description, as for swarms and flocks adopting variable behaviours in the presence of given environmental conditions. *Multiple* and *subsequent* coherent sequences of configurations corresponding to different structures are not hierarchical, but sequential and coherent over time, i.e., they display to the observer the *same* emergent, acquired property. This coherence is taken into consideration in the project mentioned in Section 7 as being properly represented by the values adopted by suitable mesoscopic variables and their properties, i.e., meta-structural properties. This is the case of the "radical" emergence of SCBs (Licata 2010b).

3.5 From structural change to meta-structural change

We consider the dynamics of SCBs as given by coherent sequences of spatial configurations adopted by interacting agents through corresponding different structures over time established by the same elements interacting in different ways, i.e., having variable structures, as introduced in Sections 3.1 and 3.2.

SCBs may indeed be considered as being established by *sequences of spatial configurations* which consist of the same elements interacting in different ways, i.e., by *sequences of variable structures* as considered for MSs and CBs. The coherence of these dynamics, defined by the emergent acquired properties, is proposed here to be suitably modelled by using Meta-Structures as mentioned above and introduced later. SCBs are not considered as *sequences of states of the same system*. Coherence between spatial configurations adopted by interacting agents through corresponding different structures over time corresponds to acquisition of emergent properties, such as shape and behaviour, and considered here as being suitably modelled by Meta-Structures and the process intended as meta-structural change.

4. New approach to model Multiple Systems and Collective Beings

In this paper and the related research project we consider a new approach to model MSs and CBs no more based on classical ergodicity, but on a more generalised way.

Ergodicity considers well defined roles and it is suitable to represent their coherence, but it does not

allow, for instance, to
- model processes characterised by dynamic changing roles, and
- set tools to induce, vary, regulate, maintain and merge emergence of processes of collective behaviours

This is the reason why we are experimenting a new approach we expected to be able to allow such a generalisation and tools.

We underline that we will maintain the approach based on considering MSs and CBs to model SCBs, but by using multiple structures corresponding to multiple systems rather than ergodicity of interchangeability of elements.

Different approaches, available in the literature will be not considered here since our project is based on reaching the above listed purposes in the limited conceptual framework of SCBs. However, as discussed later, the approach may be *empirically* generalised to collective behaviours different from the ones intended as SCBs and when modelling as MSs or CBs is not explicitly assumed, see Section 6.

## 4.1 The role of cognition

As it is well known between microscopic and macroscopic levels of description used to study Systems and Collective Behaviours we can consider the intermediate *mesoscopic* level, relating to reduced macroscopic variables without completely ignoring the degrees of freedom at the microscopic level. The adoption of a suitable mesoscopic level of description is related to the properties of the observer's cognitive system, i.e., having specific memory capabilities, image processing, cognitive processing, and input representation skills. The cognitive system presents one or more specific cognitive model(s), i.e., representations using both a level of description given, in short, by the variables, relationships with the interactions considered, scalarity, thresholds, time frame, knowledge, space within which possible actions occur, and a language representing, explaining, even abstractly, reproducing the phenomenon under consideration. A language of this kind can be related to Gestalt perception and the possibility of describing processes. Bongard proposed a method of creating an adequate language for *visual pattern recognition*, i.e., the use of a language by which the *creation* of an object could be described. His 'good continuation' principle - one of the basic principles of Gestalt psychology- assuming that the perception of a configuration includes the imaginable process of recreating, as imitating leads to the 'imitation principle', i.e., imitating the way how the adopted configuration was created, is also an effective way of describing it (Bongard 1970, Arnheim 1997). However such invention is performed by using the same cognitive system and by elaborating the models and the knowledge available. This is possible by generalising through a process of abduction, i.e., *invention of hypotheses* introduced by Pierce (Peirce 1998).

We also mention how in classical approaches based on microscopic and macroscopic variables, such as in physics and economics, single elements are considered possessing identical features and they are indistinguishable. This is a simplifying, if not reductionistic, assumption.

In this case, named *homogeneous approach*, the possibility of influencing emergent collective phenomena and their acquired properties actually focuses upon rules of interaction, energy available for the process of interaction, communications and environmental conditions. Eventual actions upon properties possessed by elements relate to their ability to react to, or elaborate interactions. However, this approach is unsuitable to model, for instance, biological systems (Pessa 2006) whose composing elements possess individual or categories of different features. In this case composing elements establish the so-called *disordered systems* for which it is theoretically impossible to model their dynamics by considering their macroscopic statistical features. An example of disordered system is given by glasses where the composing elements are arranged randomly in space vs. ordered system like crystals where the composing elements are arranged in stable patterns. A typical case of disordered system is given by *spin glass*, magnet with so-called *frustrated* *interactions* displaying stochastic disorder when ferromagnetic and anti-ferromagnetic bonding is randomly distributed like in chemical glass (Mezard *et al.* 1987, Shafee 2010). Related visual examples considering spin polarisation, i.e., alignment of elementary particles to a given direction such as in magnetic material, are presented in Figures 2 and 3.

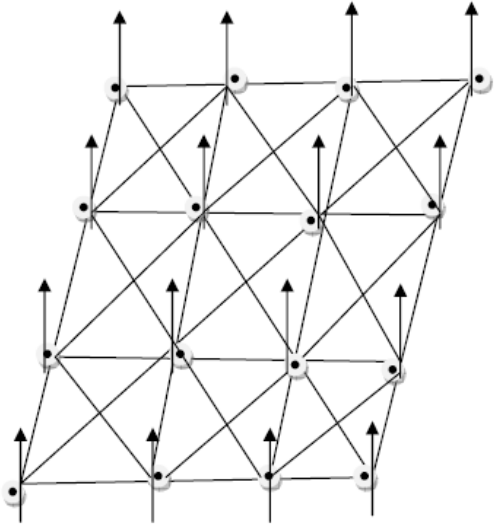

Figure 2: A visual example of ordered system

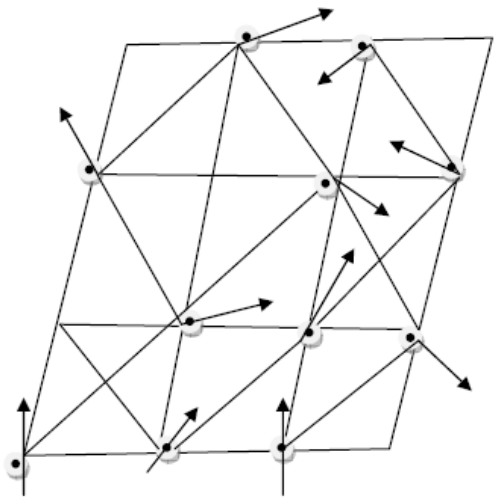

Figure 3: A visual example of disordered system

We mention as many attempts have been made to build a general theory of these systems (see, for instance, Dotsenko 1994, Chap. 8, Newman 1997, Mikhailov and Calenbuhr 2002). The problem of the relationships between the models of such systems and the emergence scenarios in QFT is still open (for a general discussion of the problem see, for instance, Pessa 2006).

Accordingly, suitable approaches to models such systems are named non-homogeneous. The approach considered here is *non-homogenous* since elements are assumed distinguishable when belonging, for instance, to different mesoscopic variables, see Section 4.2, and considering sequences of the general mesoscopic vector (5).

## 4.2 Mesoscopic state variables

The new approach considered here is based on mesoscopic state variables constructivistically identified by the observer studying a phenomenon of collective behaviour. Mesoscopic state variables represent here clusters of agents in SCB taking the same values *at the same time*.

For instance, the value adopted by a mesoscopic state variable at time $t_i$ may represent the number of elements having the same value (the observer will consider values as *equal* when *within a range of values or thresholds*) of some microscopic state variables such as the *same* distance from their nearest neighbours, the *same* speed, the *same* direction or the *same* altitude over time. However, *n*-elements constituting a mesoscopic state variable at instant $t_i$ may, in their turn, be clustered into groups having the *same* values as, in the case of distance, $n_1$ are at distance $d_1$, $n_2$ are at distance $d_2$, etc. Thus $n_1 + n_2 + ... + n_s$ may be $> n$ (same elements constitute different clusters), $< n$ (not all elements constitute different clusters) or $= n$ (each element belongs to one cluster only). It is then possible to consider the mesoscopic vectorial state variable $Va_q(t_i)$ as given by scalar values over time representing the number of elements $e_k$ at the *same* distances $d_q$: $d_1, d_2,..., d_s$ :

$$V_{dq}(t_i) = [n_1, n_2, ..., n_s].$$ (5)

Within this conceptual framework single elements may belong at time $t_i$ to one, several or no mesoscopic variables.

It is possible to generalise by substituting numbers of elements with percentages of the total number of elements.

Moreover, it is possible to consider mesoscopic state variables representing the number of elements having the same values (the observer will consider values as *equal* when *within a range of values*) of *more than one* microscopic state variable, such as the *same* distance from their nearest neighbours and the *same* speed;  the *same* speed and  the *same* direction; the *same* distance from their nearest neighbours and the *same* altitude over time; the *same* distance from their nearest neighbours, the *same* direction and the *same* altitude over time .

We then introduce the mesoscopic general vector

$$V_{k,m}(t_i) = [e_{k,1}(t_i), e_{k,2}(t_i), ..., e_{k,m}(t_i)]$$ (6)

where:

$k$  identifies one of the $k$ elements $e_k$ ;

$i$  is the computational step or instant in discretised time;

$m$ identifies one of the $m$ mesoscopic or ergodic properties  possessed by the element
$\quad e_k$ at instant $t_i$;

$e_{k,m}$ takes the  value $0$ if element $e_k$ does not have the *m-mesoscopic* or ergodic properties at time $t$;
$\quad$ or $1$ if $e_k$ does possess the *m-mesoscopic* or ergodic properties at time $t_i$.

## 4.3 Meta-elements

With reference to the mesoscopic variables introduced above we may consider Meta-elements, see Section 6.1, as time-ordered sets of values in a discrete temporal representation, for instance: values adopted by the mesoscopic variables over time; *percentages* of the total number of composing elements belonging to mesoscopic variables over time; values of thresholds  *specifying* mesoscopic state variables; values *adopted* by the mesoscopic general vector (6) introduced above; and values *adopted* by the  general index related to the *degree of respect* of the degrees of freedom for each element $e_k$, see Section 4.6. With regard to percentages mentioned above, we point out that they are related to values of *thresholds*, i.e. the larger thresholds are, the larger the number of elements belonging to mesoscopic variables is and the less significant they are. In this way variations of thresholds and related percentages are important to set the suitability to variables.

4.4 Time-ordered sets of values

Meta-elements are time-ordered sets of values in a discrete temporal representation adopted by mesoscopic variables over time and *specifying* mesoscopic state variables. While mesoscopic state variables take as values, for instance, the number(s) of elements having the same property/ies over time, Meta-elements are sets of corresponding values specifying that property, such as distance, speed, direction or altitude considered over time.

4.5 Values adopted by the mesoscopic general vector

In the same way we consider values adopted by the mesoscopic general vector $V_{k,m}(t_i)$ such as:

- How many and which elements have the same, one, several or no mesoscopic properties over time. This allows one to identify *zones* of elements possessing mesoscopic properties, their topology and dynamics.
- The total number of and which elements possess at least one mesoscopic property and the total number of and which properties are possessed by elements after the global observational computational time;
- Number of computational steps, i.e., Computational Distance (CD), occurring before *all elements* have been at least once in the *on* state (indicated as *general meso-state on*), repetitiveness;
- How many times the *general meso-state on* occurs, i.e., how many times has it taken the *on* state;
- How many and which elements possess a topological position. Topological positions may be:
- Belonging to the geometrical surface or to a specific area of interest;
- Having a specific topological distance from one of the elements such as temporary leaders and belonging to the geometrical surface or a specific area of interest;
- Be at the *topological centre* of the system, i.e., all topological distances between the element under study and all the elements belonging to the geometrical surface are equal. This element may be *virtual* and be considered as a *topological attractor* for the system. Its trajectory may *represent* the trajectory of the system.

4.6 Regularities in microscopic degrees of freedom: a virtual structure

It is possible to consider sets of values adopted over time by a general index related to the *degree of respect* of the degrees of freedom by the values adopted by variables which have to respect such degrees of freedom.
Percentages over time related to several variables such as speed and distance may be considered as representing a SCB. Properties of such percentages suitable for corresponding to and representing SCBs are studied as meta-structural properties (see Section 6.1).

5. Meta-Structures

The term *Meta-Structure* relates to simultaneous *multiple structures* over time and their sequences establishing corresponding *coherent sequences of spatial configurations*, as previously introduced in Sections 3.2, 3.3, and 3.4. We consider how they may be properly represented by mesoscopic variables and Meta-Structural properties, i.e., the properties of Meta-elements introduced above.
In this Section we list possible *meta-structural properties*, i.e., properties of values adopted by the Meta-elements (see Section 4.3).

## 5.1 Properties of ordered sets of values of Meta-elements

Meta-Structural properties are given by the mathematical properties possessed by ordered sets of values establishing Meta-elements, values adopted by mesoscopic state variables, sets of degrees of respect of the degrees of freedom and values eventually adopted by suitable macroscopic variables such as *Vol(t$_i$)*, volume of the collective entity at a given point in time and *Sur(t$_i$)*, surface of the collective entity at a given point in time.

Mathematical properties may be statistical, found through suitable statistical techniques such as Multivariate data Analysis (MDA) Cluster Analysis, Principal Component Analysis (PCA) and Principal Components (PCs), Recurrence Plot Analysis (RPA) and Recurrence Quantification Analysis (RQA); due to correlation or self-correlation; represented by interpolating functions; given by quasi-periodicity; levels of ergodicity -*degrees of ergodicity* (as introduced below in this Section), or possible relationships between them in an *N-dimensional space*, suitable for modelling a kind of *entropy of correlations*. A tool for research into meta-structural mathematical properties is the 'mesoscopic *general* vector' introduced above.

So, the approach used is based on taking into consideration mesoscopic variables, Meta-elements and Meta-Structures.

## 5.2 Degrees of ergodicity

Processes of the establishment of MSs or CBs can be suitably modelled using temporal *successions of degrees of ergodicity* in order to express the percentage of agents for which there is conceptual *interchangeability* between interacting agents which take on the *same roles at different times and different roles at the same time*, i.e., adopting an ergodic behaviour. This percentage, in turn, could be viewed as a sort of *order parameter*, suitable for describing the degree of progress of the 'phase transition' leading to the formation of the MS or CB itself. In our approach SCBs are intended as *sequences* of phase transitions, i.e., changes of configurations corresponding to different structures. The transition from the validity of a structure to another one is considered as a phase transition. The same structure may re-occur in subsequent and non-subsequent points in time.

Really, it is not only a matter of transition from the validity of a structure to that of another one but, more properly, of transitions from homogeneous areas each having a simultaneous validity of different structures per instant to other areas in the subsequent instant such that sequences of configurations are coherent.

These sequences are coherent since they establish collective entities acquiring emergent properties. This coherence is to be modelled here through meta-structural properties. Correspondingly, we conceptually consider sequences of possibly simultaneous and different values taken by mesoscopic variables and their properties represented by Meta-Structural properties. Referring now to the approach based on ergodicity when studying the coherence of MSs and CBs, we may no longer consider system monitoring to involve a single, particular, behavioural feature *F*, which will be assumed to be associated with a finite number of different possible states *F$_i$* used to detect ergodicity. We consider specific kinds of meta-structural properties as given by *multiple ergodicity* related to the ergodicity of the values adopted by different Mesoscopic state variables. The degree of ergodicity is given by:

$$E_\varphi = 1/[1 + (X_\varphi\% - Y_\varphi\%)^2]. \tag{7}$$

We may consider *Y$_\varphi$%* as the average percentage of time spent by a single element in state *F* and *X$_\varphi$% as* the average percentage of elements lying in the same state.

The state shows ergodicity when *X$_\varphi$% = Y$_\varphi$%* and the degree *E$_\varphi$* adopts its maximum value of *1*.

Different degrees of ergodicity, and their relationships, relating to different states *F$_j$* such as distance, speed and altitude are meta-structural properties and may represent coherence of a SCB.

5.3 Properties of values adopted by the mesoscopic general vector

Other kinds of meta-structural properties are given, for instance, by:

- Multiple repetitiveness, periodicity and quasi-periodicity of values adopted over time by the mesoscopic general vector $V_{k,m}(t_i)$;
- Properties of the sets of numbers of steps, elements and times as above: statistical, periodic and quasi-periodic.

6. The generalised approach

In order to introduce a suitable approach for modelling general processes of the emergence of SCBs intended as coherence between component behaviour over time, we propose to *model* properties, i.e., coherence of sequences of configurations establishing Collective Behaviour by using the properties of various, and possibly simultaneous mesoscopic variables.

The coherence of these *sequences* of configurations, i.e. their ability to establish entities acquiring emergent properties, is considered as not being reducible to structural properties such as periodicity and quasi-periodicity as assumed for processes of self-organisation, but sequences of corresponding structures.

We consider that a suitable mesoscopic level of description introduced by the observer, as in Section 4, and related meta-structural properties may be explored in the research project in order to represent multiplicity, dynamics, locality and simultaneity of sequences of different regularities we perceive in collective phenomena acquiring and maintaining emergent properties. This approach is intended as a generalisation of the one considered for MSs and CBs.

In our approach Meta-Structural properties as properties of suitable Collective variables, mesoscopic in this case, are used to model the Coherence of SCBs. The introduction of collective variables is a widely used tool in theoretical physics, allowing collective representation and for dealing with the case of so-called *quasi-particles* which share many features with traditional particles, except localization (Pessa 2009). Within this conceptual framework we consider sequences of configurations intended as being coherent when establishing entities acquiring emergent properties, as represented by suitable mesoscopic clusters of elements adopting the *same* values of some suitable microscopic state variables and related meta-structural properties over time. We consider *instantaneous, subsequent values* adopted over time by suitable mesoscopic variables, such as

- the number of elements at the *same* distance, by considering a suitable variable *distance threshold*, see Section 4;
- the properties of parameter values defining the mesoscopic variable, such as the distance considered, i.e., Meta-elements, see Section 4.3;
- mathematical properties, i.e., Meta-Structures, possessed by temporally ordered sets of values adopted over time by mesoscopic state variables and Meta-elements, see Section 5.

Within a more general framework we consider the research hypothesis that *any* sequence of *spatial configurations adopted by interacting agents through corresponding different structures over time* establish collective, i.e., coherent, behaviour when respecting suitable meta-structural properties. The process of the establishment of MSs and CBs is a particular case, consisting of sequences of systems of ergodic percentages of conceptually interchangeable agents. This case is considered as being represented in a more general way by meta-structural properties. In the Table 1 the reader can find a summary of the key concepts used in the project.

| Meta-elements, see Section 4.3, 4.4 and 6.1 | Meta-elements are time-ordered sets of values in a discrete temporal representation adopted by mesoscopic variables over time and specifying mesoscopic state variables, e.g., values of the same distance, speed or altitude adopted by elements having this mesoscopic property. |
|---|---|
| Meta-Structure, see Section 5 | The term Meta-Structure relates to simultaneous multiple structures governing interactions between elements and their sequences establishing corresponding coherent sequences of spatial configurations. |
| Meta-structural properties, see Section 5.1 | Meta-Structural properties are given by the mathematical properties possessed by ordered sets of values establishing Meta-elements, e.g., statistical, periodicity, and interpolation. |

Table 1- Summary of key concepts used in the project

## 6.1 Data for modelling

In this section we list variables whose values represent meta-structural properties, i.e., Meta-Structures, in order to allow a more concrete reasoning and present to the reader what we expect to find. Examples are given by:

a) Properties of ordered sets of values of Meta-elements

We consider values adopted over time by variables like number of elements having the maximum and minimum distance; the *same* distance from the nearest neighbour, the *same* speed, altitude and topological position at a given point in time. Macroscopic variables may be related to the measures of volume and surface of the collective entity at a given point in time. See Section 4.2. Single and coupled plotted values also related to logical combinations of mesoscopic properties over time are considered for statistical properties, interpolating functions; quasi-periodicity; levels of ergodicity.

b) Degrees of ergodicity

It is possible to consider indices of ergodicity related to properties for setting mesoscopic variables. The index of ergodicity considered over the total observation time is given by applying (7) to different cases. Consider, for instance:

1) *same* distance
   - $x_1 (t_i)$ = average percentage of elements having the *same* distance from the nearest neighbour at time $t_i$;
   - $y_1 (t_i)$ = average percentage of time spent by a single element having the *same* distance from its nearest neighbour.

2) *same* speed
   - $x_2 (t_i)$ = average percentage of elements having the *same* speed at time $t_i$;
   - $y_2 (t_i)$ = average percentage of time spent by a single element having the *same* speed.

3) *same* direction
   - $x_3 (t_i)$ = average percentage of elements having the *same* direction at time $t_i$;
   - $y_3 (t_i)$ = average percentage of time spent by a single element having the *same* direction.

4) *same* altitude
  - $x_4 (t_i)$ = average percentage of elements having the *same* altitude at time $t_i$;
  - $y_4 (t_i)$ = average percentage of time spent by a single element having the *same* altitude.

5) *same* topological position
  - $x_5 (t_i)$ = average percentage of elements having the *same* topological position at time $t_i$;
  - $y_5 (t_i)$ = average percentage of time spent by a single element having the *same* topological position.

The set of values $E_{i:1-5}$ and their relations are considered to represent possible *invariants* of the collective behaviours under study.

c) values adopted by the mesoscopic general vector

When considering the mesoscopic general vector $V_{k,m}(t_i) = [e_{k,1}(t_i) , e_{k,2}(t_i) , ..., e_{k,m}(t_i)]$ it is possible to detect properties such as:

(1) percentage over time of the total number of elements belonging to mesoscopic variables also allowing adjustment of the thresholds considered to obtain maximum significance;

(2) multiple repetitiveness, i.e., how many times a specific mesoscopic general vector $V'_{k,m}(t_i) = [e_{k,1}(t_i) , e'_{k,2}(t_i) , ..., e'_{k,m}(t_i)]$ is repeated during the total observation time and whether or not it occurs;

(3) periodicity and quasi-periodicity of values adopted over time by the mesoscopic general vector;

(4) statistical properties of sets of numbers of steps, elements and times.

d) Properties of sets of degrees of freedom

For instance, the value of the speed $Se_k (t)$ of a *k*-boid, component of a flock under study, at time *t*, must not only respect the degrees of freedom, say $S_{max}$ and $S_{min}$, but also be considered to set the degree of respect, as a percentage, of those degrees of freedom, given by

$$( [Se_k(t) - S_{min} ] * 100 ) / [S_{max} - S_{min} ]. \qquad (8)$$

With reference to distances at time *t* between two boids, say $ek_n$ and $ek_m$, we consider $[D_{max} - D_{min} ]$ and $D[ek_n-ek_m] (t)$. Percentage used by elements $ek_n$ and $ek_m$ at instant $t_i$ of their degree of freedom related to the maximum and minimum distance between them is given by

$$((D[ek_n-ek_m] (t) - D_{min} )* 100) / [D_{max} - D_{min} ] \qquad (9)$$

The properties of these time-ordered sets of values are considered to represent SCBs.

7. Purpose of the project

The purpose of the project (Meta-structures project 2009) is to:

(1) *find* Meta-Structural properties in simulated and real Collective Behaviours. We expect to find meta-structural properties as introduced above fit for eventually characterising *types* of collective behaviours such as markets, industrial districts and financial phenomena in economics, swarms, flocks and traffic in social systems, and population dynamics in biology, so as to allow the researcher, for instance, to detect the establishment and degeneration of the processes. We expect for example, to find meta-structural properties corresponding to specific collective behaviours such as types of traffic, Industrial Districts, molecular effects in biology and markets. In such a way, it should be possible to figure out typical evolutionary processes under specific environmental conditions and configurations.

(2) *prescribe* meta-structural properties to simulated and real Collective Behaviours. We expect to design suitable tools based, for instance, on changing the degrees of usage by interacting agents of the degrees of freedom in such a way as to *vary* or *prescribe* meta-structural properties to Collective interacting agents or to SCBs. Examples of areas of application are a) economics when dealing with industrial districts; b) architecture when considering how structures of space, intended as boundary conditions (Minati and Collen 2009), are able to induce the emergence of acquired properties in social habitat systems. Other possible approaches include suitably changing non-homogeneous, variable environmental properties influencing, for instance, the exchange of information (e.g., the use of noise, truncation, different representation, etc.) or usage-availability of energy. A fundamental theoretical issue relates to the *transformation* of meta-structural properties into suitable environmental properties.

## 8. The experimental activity

The experimental activity (Minati 2011) must be carried out in research domains where it is possible to have at each instant all the required information relating to elements establishing a SCB available. This is realistic, for example, for simulations and economic data for Industrial Districts depending on scalarity. Following the identification of a suitable mesoscopic level of description the researcher is expected to obtain information on labelled elements with respect to time and belonging to single mesoscopic variables allowing the measurement of data listed in Section 6.1. A first draft of a complete experimental protocol for research into meta-structural properties in simulated SCBs is under implementation. The purpose is to research meta-structural properties in simulated collective behaviours. That is, to test and experiment the approach using a simple framework, namely computational emergence where the processes of interaction occur through fixed, learning evolutionary rules. We consider finding Meta-Structures in this simplified framework as a necessary condition before the approach can be tentatively applied to non-simulated emergence such as real swarms and flocks, industrial districts, markets and traffic. Simulated collective behaviours should be at a sufficient level of complexity, for instance at Class 4 with reference to *Wolfram's classes of cellular automata*.

## 9. Future lines of research and applications

Future lines of research will attempt to:
- consider meta-structural research also in *sequences* of configurations of correlated, but even non-interacting elements such as points of sequences of images, notes in music or words in a written text. *The focus is not placed upon relations, nor upon interactions between variables, but rather upon the properties of sequences*;
- multiple meta-structural properties, *equivalent* when modelling the same collective behaviour but by using different meta-elements and mesoscopic variables;
- explore possible theoretical relationships between models of local (Shannon-Turing) and global information such as Tsallis or Fisher information (Frieden and Gatenby 2006, Tsallis 2009) and Meta-Structures. We expect, given our purely mesoscopic approach, the meta-structural properties are "invisible" to local information, but may show interesting characteristics of information which take into consideration the system as a whole;
- consider different definitions and approaches to meta-structures such as assuming the formation of meta-structures in presence of topological defects like dislocations, vortices and walls within the system, and long-range interactions, see (Pessa 2011).

There are two levels of application proposed:
- *computational*, applied, for instance, to modelling collective entities established by coherent,

i.e., acquiring emergent properties, information as for human-machine interface data, customer profiling, image recognition and image understanding where meta-structural properties may relate to categories;

- *methodological*, applied, for instance, to model collective entities established by coherent behaviours, i.e., collective behaviours such as industrial districts, markets, cities, swarm intelligence in root apex (Masi 2009), processes such as climate change, military and safety scenarios where meta-structural properties may model, in order to facilitate their *management*, induction, retention, modification of properties, and their merging.

Conclusions

We have presented here the general principles on which this approach is based including the use of variable structures, the previously introduced concepts of MSs and the consideration of SCBs as *coherent sequences* of phase transitions, i.e., coherent sequences of spatial configurations adopted by interacting agents through corresponding different structures over time.
The stability and coherence of such variable structures, i.e., the acquisition of Collective Behaviour, are assumed to be suitably represented by meta-structural properties. A possible advantage of approaches based on Meta-Structures over other available approaches used, for instance, for simulations such as Agent-based Models (ABM), Cellular Automata (CA), genetic algorithms, Neural Networks, and Game Theory models, is the expected availability of tools able to detect and manage processes of acquisition of emergent properties by acting upon very high mesoscopic degrees of freedom, without explicit invasive microscopic or macroscopic interferences. Moreover, as we have underlined many times, we aim to study situations where the dynamics are not predefined and fixed and so the mesoscopic variables are, successively, identified with relation to the observer's objectives. This kind of systems is quite typical in biological, cognitive and social sphere. Indeed, as counter-intuitive as it may sounds, when a system changes, the most important aspect is not the modification of the "microscopic" components or the modification of the global properties , but them both are re-modelled in interacting with the environment. This is the role of the configurational variables, which lead the system into its new form and mediate between the microscopic and macroscopic level, and these are just the mesoscopic variables, strongly connected to the metastable state of the change! It is easy to understand our choice of the term "meta-structures, intended as the mathematical characterization of the metastable states individuated by the observer as the expressions of the changing of a system. Meta-Structures are expected to be suitable for setting and developing behavioural strategies with systems of high mesoscopic degrees of freedom in situations of unpredictable emergence. We conclude by underlining that the theoretical novelty of the approach under study compared, for instance, to Recurrence Quantification Analysis (RQA) and classical microscopic ergodicity, lies in the emergence of mesoscopic variables related to the level of description adopted by the observer. In the above mentioned cases and, for instance, in Synergetics, the dynamics are predetermined, whereas in our approach the dynamics themselves and the variables are emergent. We considered emergence as mesoscopic coherence.